\documentclass[aps,prd,twocolumn,showpacs,groupedaddress,floatfix]{revtex4-1}

\usepackage{graphicx}
\usepackage{dcolumn}
\usepackage{bm}
\usepackage{amsmath}
\usepackage{amssymb}

\newcommand{\beq}{\begin{equation}}
\newcommand{\eeq}{\end{equation}}
\newcommand{\bea}{\begin{eqnarray}}
\newcommand{\eea}{\end{eqnarray}}

\begin{document}

\title{Flux tubes in the SU(3) vacuum: \\
London penetration depth and coherence length}

\author{Paolo Cea}
\email{paolo.cea@ba.infn.it}
\affiliation{Dipartimento di Fisica dell'Universit\`a di Bari, I-70126 Bari, 
Italy \\
and INFN - Sezione di Bari, I-70126 Bari, Italy}

\author{Leonardo Cosmai}
\email{leonardo.cosmai@ba.infn.it}
\affiliation{INFN - Sezione di Bari, I-70126 Bari, Italy}

\author{Francesca Cuteri}
\email{francesca.cuteri@cs.infn.it}
\affiliation{Dipartimento di Fisica dell'Universit\`a della Calabria,
I-87036 Arcavacata di Rende, Cosenza, Italy \\
and INFN - Gruppo collegato di Cosenza, I-87036 Arcavacata di Rende, Cosenza, 
Italy}

\author{Alessandro Papa}
\email{papa@cs.infn.it}
\affiliation{Dipartimento di Fisica dell'Universit\`a della Calabria,
I-87036 Arcavacata di Rende, Cosenza, Italy \\
and INFN - Gruppo collegato di Cosenza, I-87036 Arcavacata di Rende, Cosenza, 
Italy}

\date{\today}           
                        
\begin{abstract}
Within the dual superconductor scenario for the QCD confining vacuum, the 
chromoelectric field generated by a static $q\overline{q}$ pair can be fitted 
by a function derived, by dual analogy, from a simple variational model for 
the magnitude of the normalized order parameter of an isolated Abrikosov 
vortex. Previous results for the SU(3) vacuum are revisited,
but here the transverse chromoelectric field is measured by means of 
the connected correlator of two Polyakov loops and, in order to reduce 
noise, the smearing procedure is used instead of cooling.
The penetration and coherence lengths of the flux tube are then extracted 
from the fit and compared with previous results.
\end{abstract}

\pacs{11.15.Ha, 12.38.Aw}

\maketitle

\section{Introduction}
It is well established that in the QCD vacuum at zero temperature two static 
color charges give rise to chromoelectric flux tubes signalling color 
confinement~\cite{Bander:1980mu,Greensite:2003bk}.
As a matter of fact, Monte Carlo simulations of lattice QCD allow a 
nonperturbative study of tube-like structures that emerge by analyzing the 
chromoelectric fields between static quarks~\cite{Fukugita:1983du,
Kiskis:1984ru,Flower:1985gs,Wosiek:1987kx,DiGiacomo:1989yp,DiGiacomo:1990hc,
Singh:1993jj,Cea:1992sd,Matsubara:1993nq,Cea:1992vx,Cea:1993pi,Cea:1994ed,
Cea:1994aj,Cea:1995zt,Bali:1994de,Haymaker:2005py,D'Alessandro:2006ug}.

This suggests us a direct physical analogy between the QCD vacuum and an 
electric superconductor. Indeed, 't Hooft~\cite{'tHooft:1976ep} 
and Mandelstam~\cite{Mandelstam:1974pi} conjectured long time ago that
the vacuum of QCD could be modeled as a coherent state of color magnetic 
monopoles, namely as a dual superconductor~\cite{Ripka:2003vv}. 
In the dual superconductor model of QCD vacuum the condensation of color 
magnetic monopoles is analogous to the formation of Cooper pairs in the 
BCS theory of superconductivity. Even though the dynamical formation of 
color magnetic monopoles is not explained by the 't Hooft construction, there 
are convincing lattice evidences~\cite{Shiba:1994db,Arasaki:1996sm,Cea:2000zr,
Cea:2001an,DiGiacomo:1999fa,DiGiacomo:1999fb,Carmona:2001ja,Cea:2004ux,
D'Alessandro:2010xg} for the color magnetic condensation in the QCD 
vacuum. It should be remarked, however, that the color magnetic monopole 
condensation in the confinement mode of QCD could be a consequence, rather 
than the origin, of the mechanism of color confinement~\cite{'tHooft:2004th}.
Notwithstanding, the dual superconductivity picture of the QCD vacuum remains 
at least a useful phenomenological frame to interpret the vacuum dynamics.

In previous studies~\cite{Cea:1992vx,Cea:1993pi,Cea:1994ed,Cea:1994aj,
Cea:1995zt,Cardaci:2010tb} color flux tubes made up of chromoelectric field 
directed along the line joining a static quark-antiquark pair have
been investigated for both SU(2) and SU(3) gauge theories. In particular, 
to explore on the lattice the field configurations produced by a static 
quark-antiquark pair, the following connected correlation 
function~\cite{DiGiacomo:1989yp,DiGiacomo:1990hc,Kuzmenko:2000bq,
DiGiacomo:2000va} was used:
\begin{equation}
\label{rhoW}
\rho_W^{\rm conn} = \frac{\left\langle {\rm tr}
\left( W L U_P L^{\dagger} \right)  \right\rangle}
              { \left\langle {\rm tr} (W) \right\rangle }
 - \frac{1}{N} \,
\frac{\left\langle {\rm tr} (U_P) {\rm tr} (W)  \right\rangle}
              { \left\langle {\rm tr} (W) \right\rangle } \; ,
\end{equation}
where $U_P=U_{\mu\nu}(x)$ is the plaquette in the $(\mu,\nu)$ plane, connected
to the Wilson loop $W$ by a Schwinger line $L$, and $N$ is the number of colors
(see Fig.~1 in Refs.~\cite{Cea:1995zt,Cardaci:2010tb}).
The correlation function defined in Eq.~(\ref{rhoW}) measures the field 
strength, since in the naive continuum limit~\cite{DiGiacomo:1990hc}
\begin{equation}
\label{rhoWlimcont}
\rho_W^{\rm conn}\stackrel{a \rightarrow 0}{\longrightarrow} a^2 g 
\left[ \left\langle
F_{\mu\nu}\right\rangle_{q\bar{q}} - \left\langle F_{\mu\nu}
\right\rangle_0 \right]  \;,
\end{equation}
where $\langle\quad\rangle_{q \bar q}$ denotes the average in the presence of 
a static $q \bar q$ pair and $\langle\quad\rangle_0$ is the vacuum average. 
Accordingly, we are led to define the quark-antiquark field strength tensor as:
\begin{equation}
\label{fieldstrengthW}
F_{\mu\nu}(x) = \sqrt\frac{\beta}{2 N} \, \rho_W^{\rm conn}(x)   \; .
\end{equation}

As is well known from the usual electric superconductivity, 
tube-like structures arise as a solution of the Ginzburg-Landau 
equations~\cite{Abrikosov:1957aa}. Similar solutions were found by Nielsen and 
Olesen~\cite{Nielsen:1973cs} in the case of the Abelian Higgs model,
namely the relativistic version of the Ginzburg-Landau theory. 
In the dual superconductor model of the QCD vacuum, the formation of the 
chromoelectric flux tube can be interpreted as dual Meissner effect. In this 
context the transverse shape of the longitudinal chromoelectric field 
$E_l$ should resemble the dual version of the Abrikosov vortex field 
distribution. Therefore, the proposal was 
advanced~\cite{Cea:1992sd,Cea:1992vx,Cea:1993pi,Cea:1994ed,Cea:1994aj,
Cea:1995zt} to fit the transverse shape of the longitudinal chromoelectric 
field according to
\begin{equation}
\label{London}
E_l(x_t) = \frac{\Phi}{2 \pi} \mu^2 K_0(\mu x_t) \;,\;\;\;\;\; x_t > 0 \; ,
\end{equation}
where $K_n$ is the modified Bessel function of order $n$, $\Phi$ is
the external flux, and $\lambda=1/\mu$ is the London penetration length. 
Note that Eq.~(\ref{London}) is valid as long as $\lambda \gg \xi$, 
$\xi$ being the coherence length (type-II superconductor), which measures the 
coherence of the magnetic monopole condensate (the dual version of the Cooper 
condensate). 
However, several numerical studies~\cite{Suzuki:1988yq,Maedan:1989ju,
Singh:1993ma,Singh:1993jj,Matsubara:1994nq,Schlichter:1997hw,Bali:1997cp,
Schilling:1998gz,Gubarev:1999yp,Koma:2001ut,Koma:2003hv} in both SU(2) and 
SU(3) lattice gauge theories have shown that the confining vacuum behaves 
much like an effective dual superconductor, which lies on the borderline 
between a type-I and a type-II superconductor. If this is the case, 
Eq.~(\ref{London}) is no longer adequate to account for the transverse 
structure of the longitudinal chromoelectric field. In fact, 
in Ref.~\cite{Cea:2012qw} it has been suggested that lattice data for 
chromoelectric flux tubes can be analyzed by exploiting the results presented 
in Ref.~\cite{Clem:1975aa}, where, from the assumption of a simple variational 
model for the magnitude of the normalized order parameter of an isolated 
vortex, an analytic expression is derived for magnetic field and 
supercurrent density that solves the Ampere's law and the 
Ginzburg-Landau equations.  As a consequence, the transverse distribution of  
the chromoelectric flux tube can be described according to~\cite{Cea:2012qw}
\begin{equation}
\label{clem1}
E_l(x_t) = \frac{\phi}{2 \pi} \frac{1}{\lambda \xi_v} \frac{K_0(R/\lambda)}
{K_1(\xi_v/\lambda)} \; ,
\end{equation}
where
\begin{equation}
\label{rrr}
 R=\sqrt{x_t^2+\xi_v^2}    \;,
\end{equation}
and $\xi_v$ is a variational core-radius parameter.
Equation~(\ref{clem1}) can be rewritten as
\begin{equation}
\label{clem2}
E_l(x_t) =  \frac{\phi}{2 \pi} \frac{\mu^2}{\alpha} \frac{K_0[(\mu^2 x_t^2 
+ \alpha^2)^{1/2}]}{K_1[\alpha]} \; ,
\end{equation}
with
\begin{equation}
\label{alpha}
\mu= \frac{1}{\lambda} \,, \quad \frac{1}{\alpha} =  \frac{\lambda}{\xi_v} \,.
\end{equation}
By fitting Eq.~(\ref{clem2}) to flux-tube data, one can obtain both the 
penetration length $\lambda$ and the ratio of the penetration length to the 
variational core-radius parameter $\lambda/\xi_v$. Moreover,   
the Ginzburg-Landau $\kappa$ parameter can be obtained by
\begin{equation}
\label{landaukappa}
\kappa = \frac{\lambda}{\xi} =  \frac{\sqrt{2}}{\alpha} 
\left[ 1 - K_0^2(\alpha) / K_1^2(\alpha) \right]^{1/2} \,.
\end{equation}
Finally, the coherence length $\xi$ can be obtained by combining  
Eqs.~(\ref{alpha}) and~(\ref{landaukappa}).
Our aim is to extend previous studies of the structure of flux tubes 
performed at zero temperature to the case of SU(3) pure gauge theory at 
finite temperatures. In fact, the nonperturbative study of the chromoelectric
flux tubes generated by static color sources at finite temperature is directly 
relevant to clarify the formation of $c \bar{c}$ and $b \bar{b}$ bound states
in heavy ion collisions at high energies. To implement this program, however, 
we cannot employ the Wilson loop operator in the connected correlation 
in Eq.~(\ref{rhoW}). This problem can be easily overcome if we 
replace in Eq.~(\ref{rhoW}) the Wilson loop with two Polyakov lines.
In addition, we need to replace the cooling mechanism previously used
to enhance the signal-to-noise ratio.
Indeed, cooling is a well established method for locally 
suppressing quantum fluctuations in gauge field configurations. 
However, at finite temperatures the cooling procedure tends to 
suppress also thermal fluctuations.
Fortunately, there is an alternative, yet somewhat related, approach that
is the application of APE smearing~\cite{Falcioni1985624,Albanese1987163}
to the gauge field configurations. This approach also leads to the desirable 
effect of suppressing lattice artifacts at the scale of the cutoff
without affecting the thermal fluctuations. Moreover, this procedure can 
be iterated many times to obtain smoother and smoother gauge field 
configurations.
Obviously, we must preliminarily check that this method gives results
which are consistent with previous studies obtained with Wilson loops
and cooling. In this paper we present numerical results
on the chromoelectric flux tubes generated by static color sources in SU(3) 
pure gauge theory at zero temperature obtained with connected correlations 
built with Polyakov lines and smeared gauge links.

The plan of the paper is as follows. The connected correlation
built with Polyakov lines, used in this paper, is reported in 
Section~\ref{numerical}. In Section~\ref{su3data} we present our numerical 
results for SU(3). In Section~\ref{lengths}  we check the scaling of the 
penetration and coherence lengths and compare with previous studies. 
Finally, in Section~\ref{conclusions} we summarize our results and present our 
conclusions. 
\begin{figure}[h] 
\centering
\includegraphics[width=0.5\textwidth]{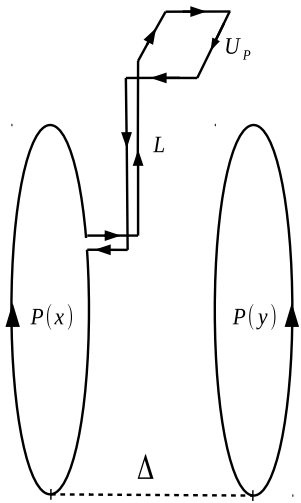} 
\caption{The connected correlator given in Eq.~(\protect\ref{eq:rhopconn})
between the plaquette $U_{P}$ and the Polyakov loops 
(subtraction in $\rho_{P}^{\rm conn}$ not explicitly drawn).}
\label{loop&field}
\end{figure}
\section{Flux tubes on the lattice}
\label{numerical}
According to our previous discussion, we shall consider the following
connected correlations (depicted in Fig.~\ref{loop&field}):  
\bea
\label{eq:rhopconn}
\rho_{P}^{\rm conn}&=&\frac{\left\langle \mathrm{tr}\left(P\left(x\right)LU_{P}
L^{\dagger}\right)\mathrm{tr}P\left(y\right)\right\rangle }{\left\langle 
\mathrm{tr}\left(P\left(x\right)\right)\mathrm{tr}\left(P\left(y\right)\right)
\right\rangle } \\
&-&\frac{1}{3}\frac{\left\langle \mathrm{tr}\left(P\left(x\right)\right)
\mathrm{tr}\left(P\left(y\right)\right)\mathrm{tr}\left(U_{P}\right)\right
\rangle }{\left\langle \mathrm{tr}\left(P\left(x\right)\right)\mathrm{tr}
\left(P\left(y\right)\right)\right\rangle}\; , \nonumber
\eea
where the two Polyakov lines separated by a distance $\Delta$ replace the 
Wilson loop in Eq.~(\ref{rhoW}). 
Taking into account Eqs.~(\ref{rhoWlimcont}) and~(\ref{fieldstrengthW}), we 
may define the field strength tensor as
\begin{equation}
\label{fieldstrengthP}
 F_{\mu\nu}\left(x\right)=\sqrt{\frac{\beta}{6}}\rho_{P}^{\rm conn}
\left(x\right).
\end{equation}
A detailed derivation of Eq.~(\ref{fieldstrengthP}), together with 
the discussion of its physical interpretation, can be found in 
Ref.~\cite{Skala:1996ar}.
\begin{figure}[htb]
\includegraphics*[width=0.95\columnwidth,clip]
{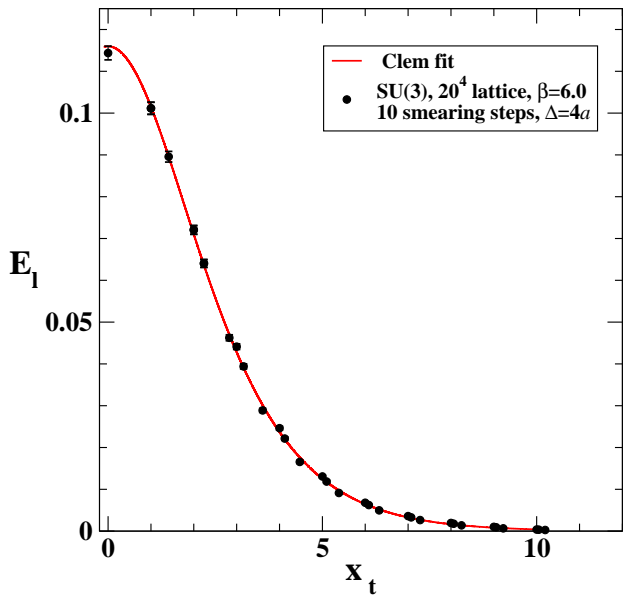} 
\caption{(color online). The longitudinal chromoelectric field 
$E_{l}$ versus $x_t$ at $\beta=6.0$ and for $\Delta=4a$, after 10 smearing 
steps. Intermediate distances are included. Full line is the best fit 
using Eq.~(\ref{clem2}).}
\label{campo}
\end{figure}
We performed numerical simulations on $20^4$ lattices using the Wilson action 
with periodic boundary conditions and the Cabibbo-Marinari 
algorithm~\cite{Cabibbo1982387} combined with overrelaxation on SU(2) 
subgroups. We considered Polyakov lines separated by $\Delta = 4a, 6a, 8a$
(where $a$ is the lattice spacing) for four different values of the gauge 
coupling $\beta$ in the range $5.9 \div 6.1$.
In order to reduce the autocorrelation time, measurements were taken after 10 
updatings. The error analysis was performed by the jackknife method over 
bins at different blocking levels.  To reduce statistical errors we 
employed the smearing procedure as described in 
Ref.~\cite{Falcioni1985624,Albanese1987163}, with smearing
parameter $\epsilon = 0.5$.  We checked that numerical results are
stable, within the statistical uncertainties, under small variations of the
parameter $\epsilon$.

As in previous studies, we confirm that the flux tube is almost completely 
formed by the longitudinal chromoelectric field $E_l$, which is constant along 
the flux axis and decreases rapidly in the transverse direction $x_t$. In 
Fig.~\ref{campo} we display the transverse distribution of the longitudinal 
chromoelectric field, measured at the middle point of the line connecting the 
static color sources, for the whole region $x_t \ge 0$.
To check rotational invariance, we considered also points 
calculated at noninteger distances.
We fitted our data to Eq.~(\ref{clem2}). The results are displayed in 
Fig.~\ref{campo}, where the full line is the curve fitting the data.
As it is evident, even in the present case Eq.~(\ref{clem2}) is able to 
reproduce accurately the transverse distribution of the longitudinal 
chromoelectric field. We also tried to restrict the fit only to
points at integer distances and obtained consistent values for the fit 
parameters. The unique observable effect was a drastic reduction of the 
reduced chi-square. Therefore, to save CPU time, we decided to perform 
measurements of the connected correlations, Eq.~(\ref{eq:rhopconn}), for 
integer transverse distances only.
\begin{table}[tb]
\begin{center} 
\caption{Summary of the fit values at $\beta=6.0$ for $\Delta=6a$.}
\label{Table:6.0_s6}
\begin{tabular}{cccccc}
\hline\hline
Smearing& $\phi$ & $\mu$ & $\lambda/\xi_\nu$ & $\kappa$ & $\chi_r^2$ \\ \hline
 16&	6.191(141) &	0.621(79) &	0.309(95) &	0.213(91) &	0.018\\
 18&	6.218(125) &	0.622(76) &	0.287(82) &	0.192(77) &	0.011\\
 20&	6.227(109) &	0.617(68) &	0.277(72) &	0.183(66) &	0.010\\
 22&	6.222(98) &	0.608(61) &	0.271(64) &	0.178(58) &	0.010\\
 24&	6.207(88) &	0.597(55) &	0.269(58) &	0.176(53) &	0.011\\
 26&	6.184(81) &	0.587(50) &	0.269(54) &	0.175(49) &	0.011\\
 28&	6.155(75) &	0.576(47) &	0.269(51) &	0.176(46) &	0.011\\
 30&	6.122(70) &	0.566(44) &	0.270(48) &	0.176(44) &	0.010\\
 32&	6.087(66) &	0.557(41) &	0.271(46) &	0.177(42) &	0.009\\
 34&	6.049(63) &	0.549(39) &	0.271(45) &	0.178(41) &	0.008\\
 36&	6.011(60) &	0.541(37) &	0.272(43) &	0.179(40) &	0.007\\
 38&	5.973(58) &	0.534(36) &	0.273(42) &	0.179(39) &	0.005\\
 40&	5.935(56) &	0.527(35) &	0.274(42) &	0.180(38) &	0.004\\
 42&	5.897(54) &	0.521(34) &	0.274(41) &	0.180(37) &	0.003\\
 44&	5.859(53) &	0.515(33) &	0.275(40) &	0.181(37) &	0.003\\
 46&	5.822(51) &	0.510(32) &	0.275(40) &	0.181(37) &	0.002\\
 48&	5.786(50) &	0.505(31) &	0.276(39) &	0.182(36) &	0.002\\
 50&	5.751(49) &	0.500(31) &	0.277(39) &	0.182(36) &	0.001\\
\hline\hline 
\end{tabular} 
\end{center}
\end{table}
\begin{figure}[htb]
\includegraphics*[width=0.95\columnwidth,clip]
{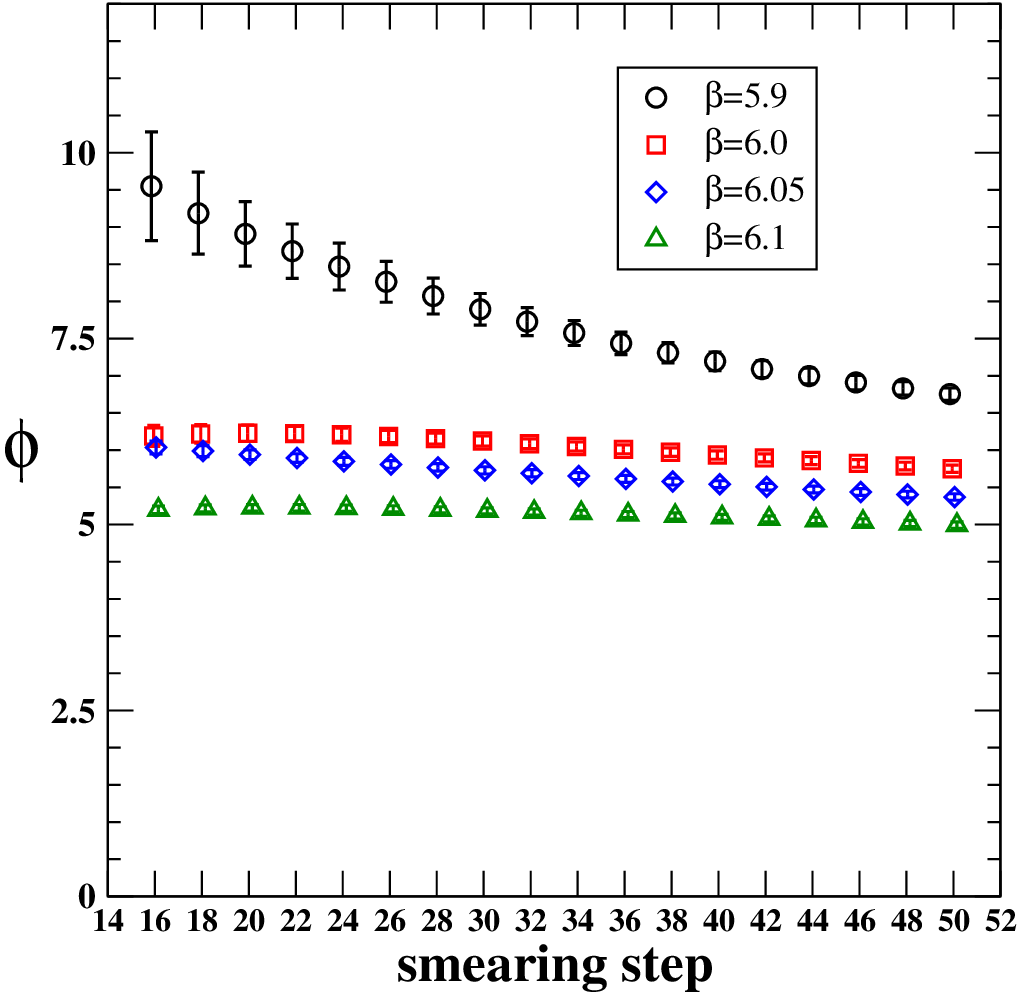} 
\caption{(color online). $\phi$ versus the smearing step for $\Delta=6a$.}
\label{phiSU3}
\end{figure}
\begin{figure}[htb]
\includegraphics*[width=0.95\columnwidth,clip]
{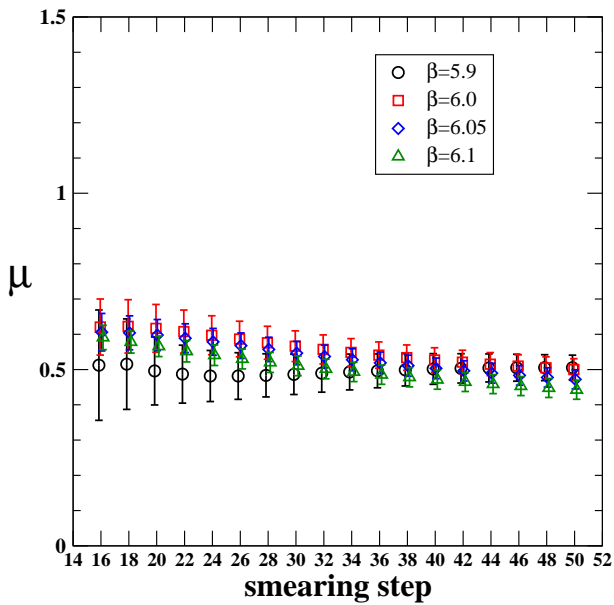} 
\caption{(color online). $\mu$ versus the smearing step for $\Delta=6a$.}
\label{muSU3}
\end{figure}
\begin{figure}[htb]
\includegraphics*[width=0.95\columnwidth,clip]
{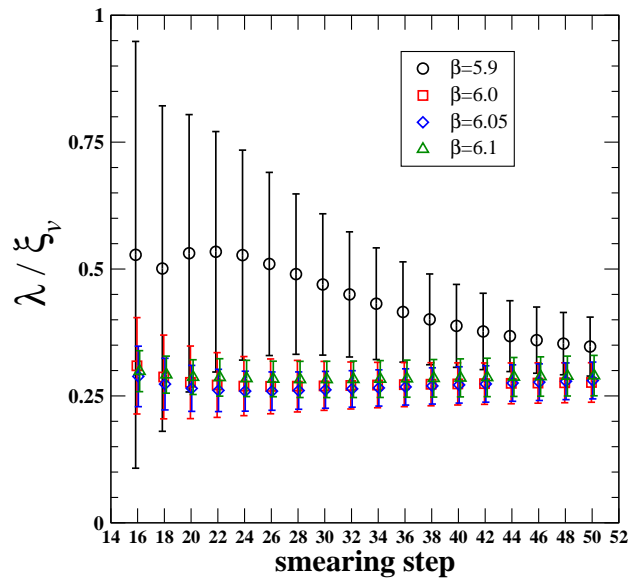} 
\caption{(color online). 
$\lambda/\xi_v$ versus the smearing step for $\Delta=6a$.}
\label{lambdasucsivSU3}
\end{figure}
\begin{figure}[htb]
\includegraphics*[width=0.95\columnwidth,clip]
{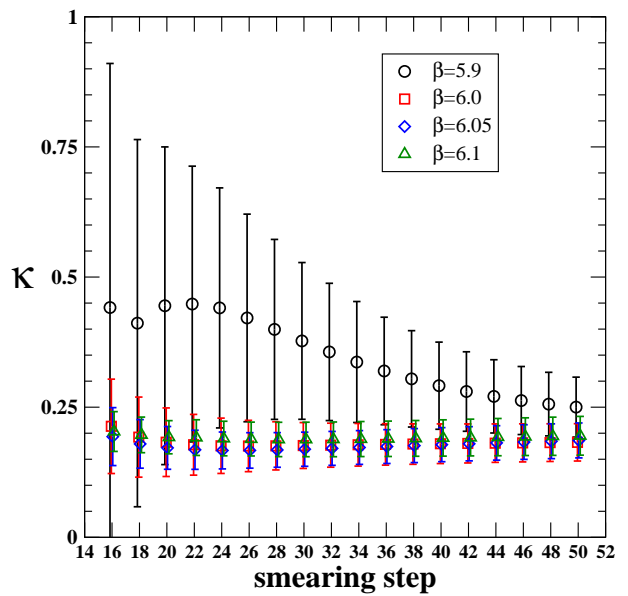} 
\caption{(color online). $\kappa$ versus the smearing step for $\Delta=6a$.}
\label{kappaSU3}
\end{figure}
%
\section{Numerical data}
\label{su3data}
We measured the connected correlator, Eq.~(\ref{eq:rhopconn}), for 
integer transverse distances $x_t$ at $\beta = 5.9, \,  6.0, \,  6.05, \, 6.1$.
To reduce statistical fluctuations in gauge field configurations
we performed measurements after several smearing steps. For each smearing,
we fitted our data for the transverse shape of the longitudinal 
chromoelectric field to Eq.~(\ref{clem2}). As a result, we obtained the fit 
parameters for different smearing steps. This allowed us to check the 
dependence of these parameters on the number of smearing steps. 
In fact, we found well defined plateaux in the extracted parameter 
values versus the smearing steps. To appreciate this point we report in 
Table~\ref{Table:6.0_s6} the values of the fit parameters for
smearing steps ranging from 16 up to 50. The parameters refer to the fit of 
the field strength tensor, corresponding to the connected correlator 
Eq.~(\ref{eq:rhopconn}) at $\beta=6.0$ and $\Delta = 6a$.
Note that the parameters $\phi$, $\mu$ and $\lambda/\xi_v$ are 
obtained by the fitting procedure, while the Ginzburg-Landau 
parameter $\kappa$ is evaluated by means of Eq.~(\ref{landaukappa}).
We looked also for contamination effects on the longitudinal 
chromoelectric field due to the presence of the static color sources. To do 
this, we varied the distance $\Delta$ between the Polyakov lines:
we found that the fitting parameters $\mu$ and $\lambda/\xi_\nu$
for $\Delta = 4a$ were systematically higher than for $\Delta = 6a, 8a$. 
On the other hand, we obtained parameters consistent within
the statistical uncertainties for the distances $\Delta = 6a$ and $8a$.
Since for $\Delta = 8a$ our estimate of the fitting parameters 
was affected by large statistical errors, we focused on the distance 
$\Delta=6a$ as a good compromise between the absence of spurious 
contamination effects due to the static color sources and a reasonable 
signal-to-noise ratio.

In Figs.~\ref{phiSU3}, \ref{muSU3}, and \ref{lambdasucsivSU3} we display the 
fitting parameters $\phi$, $\mu$ and $\lambda/\xi_v$ for different values of 
gauge coupling $\beta$ and smearing step. We see, at least for 
$\beta \ge 6.0$, that our estimate for the fitting parameters seems
to be reliable and independent of the number of smearing steps. The 
same holds also for the Ginzburg-Landau parameter $\kappa$ 
(displayed in Fig.~\ref{kappaSU3}) as obtained from Eq.~(\ref{landaukappa}).
\begin{figure}[htb]
\includegraphics*[width=0.95\columnwidth,clip]
{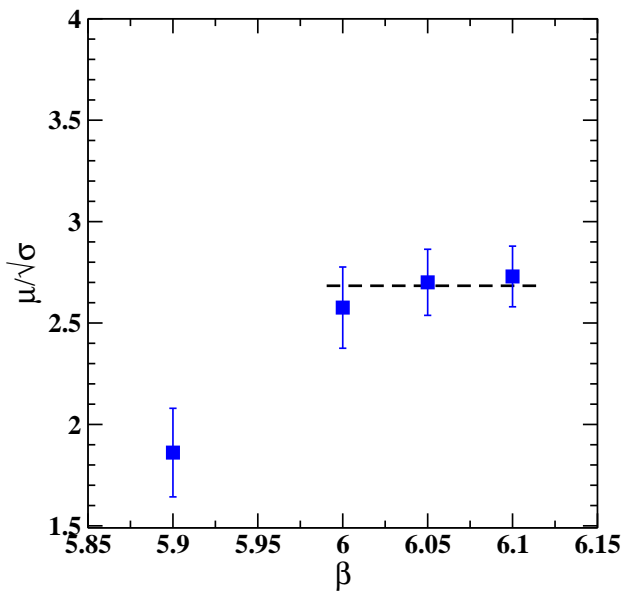} 
\caption{(color online).
$\mu/\sqrt{\sigma}$ versus $\beta$ for $\Delta=6a$.}
\label{musigmascalingsu3}
\end{figure}
\begin{figure}[htb]
\includegraphics*[width=0.95\columnwidth,clip]
{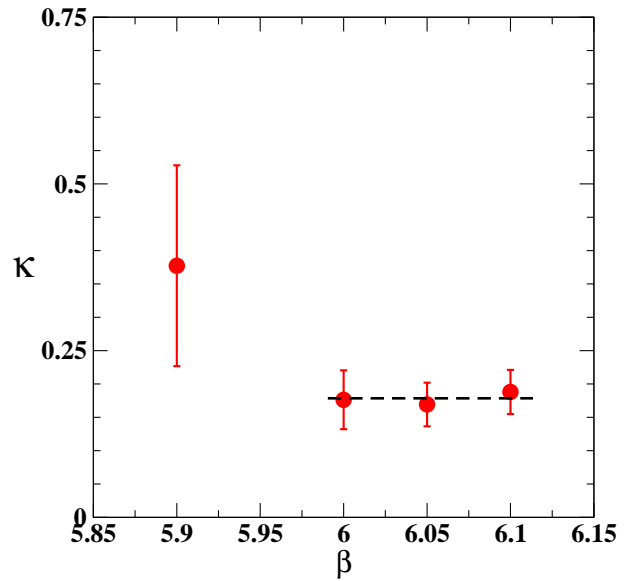} 
\caption{(color online).
$\kappa$ versus $\beta$ for $\Delta=6a$.}
\label{kappascalingsu3}
\end{figure}
\section{Penetration and coherence lengths}
\label{lengths}
Within our approach the shape of the longitudinal chromoelectric field is 
fully characterized by the London penetration depth, $\lambda$, and the 
coherence length, $\xi$.
Thus, in view of phenomenological applications in hadron physics, it is 
important to estimate these lengths in physical units. 
Firstly, we need to study the scaling of the plateau values of $a\mu$ 
with the string tension. For this purpose, we expressed these 
values of $a\mu$ in units of $\sqrt\sigma$, using the 
parameterization~\cite{Edwards:1998xf}:
\bea
\label{sqrt-sigma-SU3}
\sqrt{\sigma}(g)&=&f_{{\rm{SU(3)}}}(g^2)[1+0.2731\,\hat{a}^2(g) \\
&-&0.01545\,\hat{a}^4(g) +0.01975\,\hat{a}^6(g)]/0.01364 \;, \nonumber
\eea
\[
\hat{a}(g) = \frac{f_{{\rm{SU(3)}}}(g^2)}{f_{{\rm{SU(3)}}}(g^2(\beta=6))} 
\;, \;
\beta=\frac{6}{g^2} \,, \;\;\; 5.6 \leq \beta \leq 6.5\;,
\]
where
\beq
\label{fsun}
f_{{\rm{SU(3)}}}(g^2) = \left( {b_0 g^2}\right)^{- b_1/2b_0^2} 
\, \exp \left( - \frac{1}{2 b_0 g^2}\right) \,,
\eeq
\[
b_0 \, = \, \frac{11}{(4\pi)^2} \; \; , \; \; b_1 \, = \, \frac{102}{(4\pi)^4} 
\; .
\] \; 
In Fig.~\ref{musigmascalingsu3} we show the ratio $\mu/\sqrt{\sigma}$ for 
different values of the gauge coupling.
We see that for $\beta \ge 6.0$, $\mu$ scales according to the string 
tension. Fitting the data in the scaling window with a constant we get
\begin{equation}
\label{mu_sqrt-sigma-SU3}
\frac{\mu}{\sqrt{\sigma}} = 2.684(97) \;.
\end{equation}
Likewise, the dimensionless Ginzburg-Landau parameter $\kappa$ scales in the 
same interval of $\beta$ (see Fig.~\ref{kappascalingsu3}). Again, fitting 
with a constant gives
\begin{equation}
\label{kappa-phys}
\kappa \; = \;  0.178(21) \;.
\end{equation}
It is reassuring to see that our determinations, Eqs.~(\ref{mu_sqrt-sigma-SU3})
and~(\ref{kappa-phys}), are in good agreement with the values reported in 
Ref.~\cite{Cea:2012qw}, namely
\begin{equation}
\label{mu_sqrt-sigma-SU3-kappa}
\frac{\mu}{\sqrt{\sigma}} = 2.799(38) \;, \;\;\;\; \kappa  = 0.243(88) \;,
\end{equation}
obtained using the connected correlator built with the Wilson loop,  
Eq.~(\ref{rhoW}).
Assuming the standard value for the string tension, $\sqrt{\sigma}=420$~MeV, 
from Eq.~(\ref{mu_sqrt-sigma-SU3}) we get
\begin{equation}
\label{lambda-phys}
\lambda \; = \; \frac{1}{\mu} \;  =  \; 0.1750(63) \; {\rm fm}  \; .
\end{equation}
Combining  Eqs.~(\ref{lambda-phys}) and (\ref{kappa-phys}) we readily obtain
\begin{equation}
\label{xsi-phys}
\xi  \;  =  \; 0.983(121) \; {\rm fm}  \; .
\end{equation}
\begin{figure}[htb]
\includegraphics*[width=0.95\columnwidth,clip]
{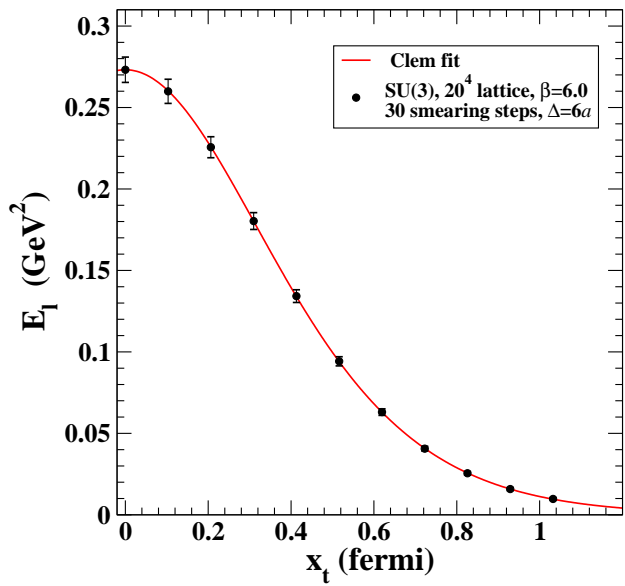}
\caption{(color online).
Longitudinal chromoelectric field $E_{l}$ versus $x_t$ in physical 
units for $\Delta=6a$ and after 30 smearing steps.
Full line is the best fit using Eq.(\ref{clem2}). }
\label{El-physical}
\end{figure}
Finally, it is interesting to display the transverse structure of the 
longitudinal chromoelectric field produced by a static quark-antiquark pair 
in physical units, see Fig.~\ref{El-physical}.
\section{Summary and Conclusions}
\label{conclusions}
In this paper we studied the chromoelectric field distribution
between a static quark-antiquark pair in the confining vacuum of the SU(3) 
pure gauge theory.

Differently from our previous studies, we adopted here a connected 
correlator built with Polyakov lines rather than Wilson loops. This is a 
preliminary and necessary step towards the extension of this analysis
to the case of nonzero temperature.
Pushing forward the dual analogy with ordinary superconductivity and 
relying on a simple variational model for the magnitude of the 
normalized order parameter of an isolated vortex, we fitted the transverse 
behavior of the longitudinal chromoelectric field according to 
Eq.~(\ref{clem2}), which allowed us to get information on the 
penetration and coherence lengths. 
We observe that what we called ``penetration length'' could match the
``intrinsic width'' of the flux tube as defined in Ref.~\cite{Caselle:2012rp},
where the adopted probe observable was the disconnected correlator of two 
Polyakov lines and a plaquette.

Our results are in good agreement with studies performed with
the connected correlator with Wilson loop, and confirm that 
the SU(3) vacuum behaves as a type-I dual superconductor. \newline
This conclusion is shared with Ref.~\cite{Shibata:2012ae}, where
the non-Abelian dual Meissner effect is investigated within the so-called 
``restricted field dominance''. More recently, the same 
authors~\cite{Shibata:2014spa,Shibata:2014tpa} presented some preliminary
studies at nonzero temperature.
Finally, we observe that our estimate of the London penetration length is 
in good agreement with the recent determination in Ref.~\cite{Bicudo:2014wka}, 
obtained using correlators of plaquette and Wilson loop not 
connected by the Schwinger line, thus leading to the (more noisy) squared 
chromoelectric and chromomagnetic fields.
\section*{Acknowledgments}
Simulations have been performed on the BC$^2$S cluster in Bari.

%

%

\end{document}